\documentclass[11pt,a4paper]{article}
\usepackage{epic,eepic,amsmath,latexsym,fullpage,amssymb,color,amsthm}
\usepackage{ifthen,graphics,epsfig}
\usepackage{times}
\usepackage[english]{babel}
\usepackage{xspace}

\usepackage{float}
\newfloat{algorithm}{thp}{lop}
\floatname{algorithm}{Algorithm}
\begin{document}
\newlength {\squarewidth}
\renewenvironment {square}
{
\setlength {\squarewidth} {\linewidth}
\addtolength {\squarewidth} {-12pt}
\renewcommand{\baselinestretch}{0.75} \footnotesize
\begin {center}
\begin {tabular} {|c|} \hline
\begin {minipage} {\squarewidth}
\medskip
}{
\end {minipage}
\\ \hline
\end{tabular}
\end{center}
}

\newcommand{\Xomit}[1]{}
\newcommand{\Red}[1]{#1}
\def\vval{{{\#}_{v}}}

\newtheorem{definition}{Definition}
\newtheorem{theorem}{Theorem}
\newtheorem{lemma}{Lemma}
\newtheorem{corollary}{Corollary}
\newtheorem{property}{Property}
\newcommand{\toto}{xxx}
\newenvironment{proofT}{\noindent{\bf
Proof }} {\hspace*{\fill}$\Box_{Theorem~\ref{\toto}}$\par\vspace{3mm}}
\newenvironment{proofL}{\noindent{\bf
Proof }} {\hspace*{\fill}$\Box_{Lemma~\ref{\toto}}$\par\vspace{3mm}}
\newenvironment{proofC}{\noindent{\bf
Proof }} {\hspace*{\fill}$\Box_{Corollary~\ref{\toto}}$\par\vspace{3mm}}
\newenvironment{proofP}{\noindent{\bf
Proof }} {\hspace*{\fill}$\Box_{Property~\ref{\toto}}$\par\vspace{3mm}}

\newcounter{linecounter}
\newcommand{\linenumbering}{\ifthenelse{\value{linecounter}<10}{(\arabic{linecounter})}{(\arabic{linecounter})}}
\renewcommand{\line}[1]{\refstepcounter{linecounter}\label{#1}\linenumbering}
\newcommand{\resetline}[1]{\setcounter{linecounter}{0}#1}
\renewcommand{\thelinecounter}{\ifnum \value{linecounter} > 9\else \fi \arabic{linecounter}}

\newenvironment{lemma-repeat}[1]{\begin{trivlist}
\item[\hspace{\labelsep}{\bf\noindent Lemma~\ref{#1} }]}%
{\end{trivlist}}

\newenvironment{theorem-repeat}[1]{\begin{trivlist}
\item[\hspace{\labelsep}{\bf\noindent Theorem~\ref{#1} }]}%
{\end{trivlist}}

\title{\bf Trading off  $t$-Resilience for Efficiency  \\
         in Asynchronous Byzantine Reliable Broadcast}

\author{Damien Imbs$^\dag$ ~~and ~~Michel Raynal$^{\star,\ddag}$ \\~\\
$^{\dag}$  LIF, Universit\'e d'Aix-Marseille, 13288 Marseille, France\\
$^{\star}$  Institut Universitaire de France\\
$^{\ddag}$  IRISA, Universit\'e de Rennes, 35042 Rennes Cedex, France \\
{\small {\tt damien.imbs@lif.univ-mrs.fr~~~~~raynal@irisa.fr}}
}
\date{}
\maketitle


\begin{abstract}
This paper presents a simple and efficient reliable broadcast 
algorithm for asynchronous message-passing systems made up of $n$ processes, 
among which up to $t<n/5$  may behave arbitrarily (Byzantine processes). 
This algorithm requires two communication steps and $n^2-1$ messages.
When compared to Bracha's algorithm, which is resilience optimal ($t<n/3$)
and requires three communication steps and $2n^2-n-1$ messages, 
the proposed  algorithm shows an interesting tradeoff between 
communication efficiency and $t$-resilience.

~~\\
{\bf Keywords}: 
Algorithm, Asynchronous system, Byzantine process, 
Distributed computing, Fault-tolerance, Message-passing, Reliable broadcast. 
\end{abstract}

\section{Introduction}

\paragraph{On reliable broadcast}
{\it Reliable broadcast} (RB) is a communication abstraction central to 
fault-tolerant distributed systems. 
It allows each process to broadcast messages to all  processes
despite failures. More precisely, it guarantees that  the non-faulty 
processes deliver the same set of messages and this set includes
at least all the messages they  broadcast. It can also contain  
messages broadcast by faulty processes. 

The fundamental property of reliable broadcast lies in the fact that 
no two correct processes deliver different sets of messages. 
This communication abstraction is a basic building block used to
build a {\it total order reliable broadcast} abstraction 
(sometimes called ``atomic broadcast''),
which adds the total order property on message delivery
(see e.g.,~\cite{AW04,B12,CGR11,HT94,L98,L96,R10}).  In turn, 
total order  broadcast is a  basic building block for state machine 
replication, which is a fundamental paradigm in fault-tolerance.  

\paragraph{Reliable broadcast in the presence of Byzantine processes}
Reliable broadcast has been studied in the context of Byzantine failures
since the eighties. A process commits a Byzantine failure if 
it behaves arbitrarily~\cite{LSP82,PSL80}. Such a behavior can be intentional 
(also called malicious) or the result of a transient fault which altered 
the content of local variables of a process, thereby modifying its intended
behavior in an unpredictable way. 

An elegant algorithm, due to G. Bracha,  implementing the  reliable broadcast
abstraction in an asynchronous system of $n$ processes which communicate by 
message-passing, and  where up to $t<n/3$ processes may be Byzantine is 
described in~\cite{B87}. This algorithm is signature-free. 
It is shown in~\cite{B87,BT85} that  $t<n/3$ is an upper bound on the number 
of Byzantine processes that can be tolerated. Hence, 
Bracha's algorithm is $t$-resilience optimal. This algorithm is based on 
a ``double echo'' mechanism of the value broadcast by the sender process. 
It uses three types of messages, requires three consecutive communication 
steps, and $(n-1)+2n(n-1)=2n^2-n-1$ underlying messages.

\paragraph{Content of the paper}
This paper presents a new signature-free Byzantine-tolerant reliable broadcast 
algorithm, which uses only two message types, two consecutive communication 
steps, and $(n-1)+n(n-1)=n^2-1$ underlying  messages. 
This gain, with respect  Bracha's algorithm, in both the time and the number 
of messages,  is obtained with a weaker $t$-resilience requirement, namely $t<n/5$ 
instead of $t<n/3$. This shows an interesting tradeoff between 
communication cost (number of communication steps\footnote{The number 
of different message types is always the same as the number of 
communication steps. This is needed to associate the appropriate 
processing to each message.} and the number of messages) on one side,  and 
fault-resilience on the other side (see Table~\ref{comparison}).

\begin{table}[ht]
\begin{center}
\renewcommand{\baselinestretch}{1}
\small
\begin{tabular}{|c||c|c|c|}
\hline
   & fault  &  communication steps & number of \\

  & resilience  &  message types  &  messages \\
\hline
Bracha's algorithm~\cite{B87} & $n>3t$ &   $3$  &  $2n^2-n-1$    \\
\hline
This paper                   &  $n>5t$ &   $2$  &  $n^2-1$ \\

\hline 
\end{tabular}
\end{center}
\vspace{-0.4cm}
\caption{Bracha's algorithm with respect to the proposed algorithm}
\label{comparison} 
\end{table}

\section{Computation Model}
\label{sec:computation-model} 

\paragraph{Asynchronous processes}
The system is made up of a finite set $\Pi$ of $n>1$ asynchronous sequential
processes, namely $\Pi=\{p_1,\ldots,p_n\}$. ``Asynchronous'' means that 
each process proceeds at its own speed, which can vary arbitrarily with 
time, and always remains  unknown to the other processes.

\paragraph{Communication network}
The processes  communicate by exchanging messages through an asynchronous 
reliable  point-to-point network. ``Asynchronous'' means that a message that 
has  been sent  is eventually  received by  its destination  process, i.e.,
there is  
no bound on message transfer delays. ``Reliable'' means that the network does 
not loose, duplicate, modify, or create messages. ``Point-to-point'' means that 
there  is  a bi-directional  communication  channel  between  each pair  of
processes.

A process $p_i$ sends a message to a process $p_j$ by invoking the primitive 
``${\sf send}$ {\sc tag}$(m)$ ${\sf to}~p_j$'', where {\sc tag} is the type
of the message and $m$ its content. To simplify the presentation, it is
assumed that a process can send messages to itself. A process receives 
a message by executing the primitive ``${\sf receive} ()$''.
The macro-operation  ``${\sf broadcast}$  {\sc tag}$(m)$'' is a shortcut for
``{\bf for} $j\in\{1,\cdots,n\}$ {\bf do} 
${\sf send}$ {\sc tag}$(m)$ ${\sf to}~p_j$ {\bf end for}''.

\paragraph{Failure model}
Up to $t$ processes can exhibit a {\it Byzantine} behavior.
 A Byzantine process is a process that behaves
arbitrarily: it can crash, fail to send or receive messages, send
arbitrary messages, start in an arbitrary state, perform arbitrary state
transitions, etc. As a simple example, a Byzantine process, 
which is assumed to send a
message $m$ to all the processes, can send a message $m_1$ to some processes, 
a different message $m_2$  to another subset of processes, and no message 
at all to the other processes. Moreover, Byzantine processes can collude 
to ``pollute'' the computation. They can also control the network in the sense 
that they can re-order message deliveries at correct processes. 
It is however assumed that a Byzantine process cannot send an infinite 
number of messages. 

Let us notice that, as each pair of processes is connected by a channel, 
a process  can identify the sender of each message it receives. 
Hence,  no  Byzantine  process  can  impersonate  another  process.   
As in Bracha's algorithm, 
this allows the proposed algorithm to be signature-free. 

A process that exhibits a Byzantine behavior is also called {\it faulty}.
Otherwise, it is {\it correct} or {\it non-faulty}.

\section{Reliable Broadcast}
\label{sec:RB-definition}
The reliable broadcast (denoted RB-broadcast) communication  abstraction  
provides each process with two operations, denoted ${\sf RB\_broadcast}()$ 
and ${\sf RB\_deliver}()$. As in~\cite{HT94}, we use the following terminology:
when a process invokes ${\sf RB\_broadcast}()$, we say that it ``RB-broadcasts
a message'', and  when it executes  ${\sf RB\_deliver}()$, we say that it 
``RB-delivers a message''. RB-broadcast is defined by the following properties. 

\begin{itemize}
\vspace{-0.2cm}
\item RB-Validity. 
If a correct process RB-delivers the message {\sc msg}$(v)$ from a correct 
process $p_i$, then $p_i$ RB-broadcast {\sc msg}$(v)$. 
\vspace{-0.2cm}
\item RB-Integrity.
A correct process RB-delivers at most one message from any process $p_i$. 
\vspace{-0.2cm}
\item 
RB-Agreement.
No two correct processes RB-deliver distinct messages from the same process.
\vspace{-0.2cm}
\item RB-Termination-1. If a correct process RB-broadcast a message,
all correct processes eventually RB-deliver this  message.
\vspace{-0.2cm}
\item RB-Termination-2.
 If a  correct process RB-delivers a message $m$  from $p_i$ 
(possibly faulty)  then all correct processes  eventually  RB-deliver 
$m$ from $p_i$. 
\end{itemize}

\paragraph{On the safety properties' side}
RB-validity  relates the output (messages RB-delivered) to the inputs 
(messages RB-broadcast). RB-integrity states that there is no duplication. 
 RB-agreement states that there is no duplicity: be the sender correct or not, 
it is not possible for a correct process to RB-deliver $m$ while 
another correct process  RB-delivers $m'\neq m$. 

\paragraph{On the liveness properties' side}
The RB-Termination properties state the guarantees on message RB-delivery. 
 RB-Termination-1 states that a message RB-broadcast by a correct process
is RB-delivered by all correct processes. 
RB-Termination-2 gives its name to reliable broadcast. Be the sender correct 
or not, every message RB-delivered by a correct process is  RB-delivered 
by all correct processes. 

It follows that all correct processes RB-deliver the same set of messages, 
and this set contains at least all the messages RB-broadcast by the 
correct processes.

\paragraph{RB-broadcasting a sequence of messages}
The previous definition considers that each correct process RB-broadcasts 
at most one message. It is easily possible to extend it to the case 
where a correct process RB-broadcasts a  sequence of messages. 
In the algorithm that follows, the identity $j$ of the sender $p_j$ must then 
be replaced by a pair $\langle j, sn\rangle$, where $sn$ is the sequence
number associated by $p_j$ with the message.

\section{A Communication-Efficient Reliable Broadcast Algorithm for $t<n/5$}

\paragraph{The algorithm}
Algorithm ~\ref{algo-RB-broadcast}, which implements the reliable broadcast 
abstraction, consists of a client side and a server side. 
On the client side, when a (correct) process wants to RB-broadcast  an
application message {\sc msg}$(v_i)$, it simply broadcasts the algorithm 
message  {\sc init}$(i,v_i)$.

On the server side, a process can receive two types of messages. 
\begin{itemize}
\vspace{-0.2cm}
\item
When it receives a message  {\sc init}$(j,v)$ (necessarily from process $p_j$
as the processes are connected by bidirectional channels), a process $p_i$ 
broadcasts the message {\sc witness}$(j,v)$ (line~\ref{BYZ-RBC-02}) if (a)
this message is the first message {\sc init}$()$ $p_i$ receives from $p_j$, and 
(b) $p_i$ has not yet broadcast a message {\sc witness}$(j,-)$.
\vspace{-0.2cm}
\item
When a process $p_i$ receives a message {\sc witness}$(j,v)$ (from any process),
it does the following.
\begin{itemize}
\vspace{-0.2cm}
\item If $p_i$ has received the same message from ``enough-1''  processes
(where ``enough-1'' is  $(n-2t)$, i.e.,  
at least $n-3t\geq 2t+1$ correct processes sent this message, 
and $p_i$ has not yet broadcast the same message {\sc witness}$(j,v)$, it 
forwards it to all processes.  This concludes the ``forwarding phase'' of 
$p_i$ as far as a message of $p_j$  is concerned. 
\vspace{-0.1cm}
\item  If $p_i$ received the same message from ``enough-2''
processes (where ``enough-2'' means ``at least $(n-t)$ 
processes'', i.e., the message was received from at least $n-2t\geq 3t+1$ 
correct processes, $p_i$ locally RB-delivers {\sc msg}$(j,v)$ if 
not yet done. This concludes the ``RB-delivering phase'' of a message 
from $p_j$,  as far as $p_i$ is concerned. 
\end{itemize} 
\end{itemize}

\begin{algorithm}[ht]
\centering{
\fbox{
\begin{minipage}[t]{150mm}
\footnotesize
\renewcommand{\baselinestretch}{2.5}
\resetline
\begin{tabbing}
aA\=aaA\=aaaaaaaaaaaaA\kill

{\bf opera}\={\bf tion} ${\sf RB\_broadcast}$ {\sc msg}$(v_i)$ {\bf is}\\

\line{BYZ-RBC-01} \>  ${\sf broadcast}$  {\sc init}$(i,v_i)$.\\~\\

{\bf when}  {\sc init}$(j,v)$  {\bf is  received from} $p_j$ {\bf  do}\\

\line{BYZ-RBC-02} \> 
     {\bf if} \big(first  reception of  {\sc init}$(j,-)$
                   and {\sc witness}$(j,-)$ not yet broadcast\big)
     {\bf then} ${\sf broadcast}$  {\sc witness}$(j,v)$
     {\bf end if}.\\~\\

{\bf when} {\sc witness}$(j,v)$  {\bf is  received do}\\
\line{BYZ-RBC-03} \>   
{\bf if} \=
   \big({\sc witness}$(j,v)$  received from $(n-2t)$
        different processes and {\sc witness}$(j,v)$ not yet broadcast\big)\\

\line{BYZ-RBC-04} \> \>   
{\bf then}  ${\sf  broadcast}$ {\sc witness}$(j,v)$\\

\line{BYZ-RBC-05} \>  {\bf end if};\\

\line{BYZ-RBC-06} \>   
   {\bf if} \=
   \big({\sc witness}$(j,v)$  received from  $(n-t)$
        different processes and {\sc msg}$(j,-)$ not yet  RB\_delivered\big)\\

\line{BYZ-RBC-07} \> \>   
{\bf then}  ${\sf RB\_deliver}$ {\sc msg}$(j,v)$\\

\line{BYZ-RBC-08} \>  {\bf end if}.

\end{tabbing}
\normalsize
\end{minipage}
}
\caption{Communication-efficient Byzantine reliable broadcast algorithm 
         for $t<n/5$}
\label{algo-RB-broadcast} 
}
\end{algorithm}

\paragraph{Cost of the algorithm}
Only two types of messages are used ({\sc init} and {\sc witness}). 
It is easy to see that the broadcast of a message by a correct process
requires two consecutive communication steps (broadcast of an 
{\sc init} message whose receptions entail at most $n$ broadcasts of  
{\sc witness} messages).
Not counting the messages that a process sends to itself, 
a reliable broadcast  by a correct process costs 
$(n-1)$ messages {\sc init} and at most $n(n-1)$ messages {\sc witness} 
(counting only the  messages under the control of the algorithm).

\section{Proof of the Algorithm}
The proof assumes $t<n/5$. 

\begin{lemma}
\label{lemma:no-Byzantine-val}
Let {\sc init}~$(i,v)$ be a message that is never broadcast by a correct 
process $p_i$. If Byzantine processes broadcast the message 
{\sc witness}~$(i,v)$, no correct process will forward this message at 
line~{\em\ref{BYZ-RBC-04}}.
\end{lemma}

\begin{proofL}
Let us consider the worst case where  $t$ processes are Byzantine
and each of them broadcasts the same message {\sc witness}~$(i,v)$.
For a correct process $p_j$ to  forward this message
at line~\ref{BYZ-RBC-04}, the forwarding predicate of line~\ref{BYZ-RBC-03}
must be satisfied. But, in order for this predicate to be true at a 
correct process $p_j$, 
this process must receive  the message {\sc witness}~$(i,v)$ 
from $n-2t$ different processes. As $n-2t>t$, this cannot occur.
\renewcommand{\toto}{lemma:no-Byzantine-val}
\end{proofL}

\begin{theorem}
  \label{theo-RB-algorithm}
Algorithm~{\em\ref{algo-RB-broadcast}} implements the reliable broadcast 
abstraction in $n$-process asynchronous message-passing systems in which
up to $t<n/5$ processes may commit Byzantine failures. 
\end{theorem}

\begin{proofT}~\\
\noindent
Proof of the RB-Validity property. \\ 
Let $p_i$ be a correct process
that invokes ${\sf RB\_broadcast}$ {\sc msg}$(v)$ and consequently
broadcasts the message {\sc init}$(i,v)$ at line~\ref{BYZ-RBC-01}.
The fact that no correct process RB-delivers a message different from
{\sc msg}$(i,v)$ comes from the following observation.  To RB-deliver
a message {\sc msg}$(i,v')$, a correct process must receive the
message {\sc witness}$(i,v')$ from more than $(n-t)$ different
processes (line~\ref{BYZ-RBC-06}).  But if the (at most) $t$ Byzantine
processes forge a fake message {\sc witness}$(i,v')$, with $v\neq v'$,
this message will never be forwarded by correct processes
(Lemma~\ref{lemma:no-Byzantine-val}). As $n-t>t$, it follows from the
predicate of line~\ref{BYZ-RBC-06} that the content of the message
RB-delivered by any correct process cannot be different from
$(i,v)$. \\

\noindent
Proof of the RB-Integrity property. \\
This property follows directly from the RB-delivery predicate of 
line~\ref{BYZ-RBC-06}, namely, at most one message {\sc msg}$(j,v)$  
can be delivered by any correct process $p_i$. \\

\noindent
Proof of the RB-Agreement property. \\ 
Let $p_k$ be a process that sends at least one message {\sc init}$(k,-)$.
If $p_k$  is correct, it sends at most one such message. If it is Byzantine, 
it may send more. Hence, let us assume that $p_k$ sends {\sc init}$(k,v_1)$, 
{\sc init}$(k,v_2)$, etc., {\sc init}$(k,v_m)$, where $m \geq 1$. 
For any $x\in[1..m]$, let $Q_x$ be the set of correct processes
that receive  the message  {\sc init}$(k,v_x)$, and these receptions directed 
them  to broadcast the message {\sc witness}$(k,v_x)$ at line~\ref{BYZ-RBC-02}. 
Due to the fact that only $p_k$ can send messages {\sc init}$(k,-)$, 
it follows from the reception predicate of line~\ref{BYZ-RBC-02} that 
a correct process can belong to at most one set $Q_x$. Hence, we have:
$(x\neq y) \Rightarrow Q_x\cap Q_y=\emptyset$. 
We consider two cases according to the size of the sets $Q_x$.  
\begin{itemize}
\vspace{-0.2cm}
\item 
Let us first consider a set $Q_x$ such that $|Q_x|<n-3t$. 
Let $p_j$ be any correct process that does not belong to $Q_x$ (hence $p_j$ 
does not process the message {\sc init}$(k,v_x)$  at line~\ref{BYZ-RBC-02}
if it receives it). As $n-t>n-3t$, $p_j$ does exist.
Process $p_j$ can receive the message {\sc witness}$(k,v_x)$
(a) from  each process of $Q_x$, 
and (b) from each of the $t$ Byzantine processes. It follows that
$p_j$ can receive {\sc witness}$(k,v_x)$ from at most $t+|Q_x|$ different 
processes.  As  $t+|Q_x|< n-2t$, the predicate of line~\ref{BYZ-RBC-03}
cannot be satisfied at $p_j$, and consequently, $p_j$ (i.e., any correct 
process $\notin Q_x$) will never send  the message {\sc witness}$(k,v_x)$. 
Hence the number of messages {\sc witness}$(k,v_x)$ received by 
any correct process can never attain $(n-t)$, from which we conclude 
that no correct process RB-delivers {\sc msg}$(k,v_x)$. 
It follows that, if there is a single set (of correct processes) 
$Q_z$ (i.e., $z=m=1)$, and this set is  such that $|Q_z|\geq n-3t$, 
at most one message {\sc msg}$(k,-)$ may be  RB-delivered by
a correct process, and this message is then {\sc msg}$(k,v_z)$. 
\vspace{-0.2cm}
\item 
Let us now consider the case where there are at least two different sets 
of correct processes $Q_x$ and $Q_y$, each of size at least $( n-3t)$. 
Let us remind that, in the worst case, each of the $t$ Byzantine processes can 
systematically play double game by sending both  {\sc witness}$(k,v_x)$ and  
{\sc witness}$(k,v_y)$ to each correct process without having received the 
associated message  {\sc init}$(k,-)$).  Moreover,  in the worst case, 
we have exactly $(n-t)$ correct processes. 
(If, in a given execution, strictly less than $t$ processes are Byzantine, 
we consider the equivalent execution in which exactly $t$ processes are 
Byzantine, and some of them behave like correct processes.)
As  both $Q_x$ and $Q_y$ contain
only correct processes, and  $Q_x\cap Q_y=\emptyset$, it follows that 
$|Q_x|+|Q_y|+t\leq n$, which implies $2n-6t+t\leq |Q_x|+|Q_y|+t\leq n$, 
from which we obtain 
$5t\geq n$, which  contradicts the assumption on $t$ (namely, $n>5t$).  
Consequently,  at least one of $Q_x$ and $Q_y$ is composed of less than 
$(n-3t)$ correct processes. It follows from the previous paragraph that
the corresponding message {\sc msg}$(k,-)$ cannot be RB-delivered by a
correct process. As this is true for any pair of  sets $Q_x$ and $Q_y$, 
it follows that,  if $p_k$ sends several messages  {\sc init}$(k,v_1)$, 
{\sc init}$(k,v_2)$, etc., {\sc init}$(k,v_m)$, at most one of them can 
give rise to a set $Q_x$ such that $|Q_x|\geq n-3t$, and, consequently, at most 
one message {\sc msg}$(k,v_x)$ can be RB-delivered by any correct process.
\end{itemize}

\noindent
Proof of the RB-Termination-1 property.\\
Let $p_i$ be a correct process that invokes 
${\sf RB\_broadcast}$ {\sc msg}$(v)$ and consequently broadcasts the message 
{\sc init}$(i,v_i)$ at line~\ref{BYZ-RBC-01}.
It follows that any correct process $p_j$ receives this message.  
Let us remember that, due to the network connectivity assumption, 
there is a channel  connecting $p_i$ to $p_j$ and consequently the message 
{\sc init}$(i,v)$ cannot be a fake message forged by a Byzantine process.
Moreover, due to Lemma~\ref{lemma:no-Byzantine-val}, no message 
{\sc witness}$(i,v')$, with $v'\neq v$, forged by Byzantine processes, can be 
forwarded by a correct process at lines~\ref{BYZ-RBC-03}-\ref{BYZ-RBC-04}.
Hence, when $p_j$ receives {\sc init}$(i,v)$, it broadcasts  
the message {\sc witness}$(i,v)$  at line~\ref{BYZ-RBC-02}.
It follows that every correct process eventually receives this message from 
$(n-t)$ different processes and consequently locally RB-delivers 
the message  {\sc msg}$(i,v)$ at lines~\ref{BYZ-RBC-06}-\ref{BYZ-RBC-08}, 
which proves the property. \\

\noindent
Proof of the RB-Termination-2 property.\\
Let $p_i$ be a correct process that RB-delivers the message {\sc msg}$(k,v)$. 
It follows that the   RB-delivery predicate of line~\ref{BYZ-RBC-06}
is true at $p_i$, and consequently, $p_i$ received the message 
{\sc witness}$(k,v)$ from at least $(n-t)$ different processes, 
i.e., from at least $n-2t>t$ correct processes. 
 
It follows that at least $(n-2t)$ correct processes broadcast 
{\sc  witness}$(k,v)$, and consequently the predicate of
line~\ref{BYZ-RBC-03} is eventually true at each correct
process. Hence, every correct process eventually broadcasts the
message {\sc witness}$(k,v)$ at line~\ref{BYZ-RBC-04}, if not yet done
before (at line~\ref{BYZ-RBC-02} or line~\ref{BYZ-RBC-04}).
As there are at least $(n-t)$ correct processes, each of them
eventually receives {\sc witness}$(k,v)$ from $(n-t)$ 
different processes, and consequently RB-delivers {\sc msg}$(k,v)$
at line~\ref{BYZ-RBC-07}, which proves the property. 
\renewcommand{\toto}{theo-RB-algorithm}

\end{proofT}


\vspace{-0.3cm}
\section*{Acknowledgments}
The authors want to thank the reviewers for their constructive
comments.  This work has been partially supported by the Franco-German
DFG-ANR Project 40300781 DISCMAT devoted to connections between
mathematics and distributed computing, and the French ANR project
DESCARTES devoted to distributed software engineering.

\vspace{-0.1cm}


\end{document}